# Perfumer: A Large Language Model-based Agentic Framework Accelerating Structural Relaxation and Electronic Band Prediction for First-Principle Calculation

*Zhiwei Li*[1,†], *Yuhao Zhou*[2,†], *Yikai Yang*[1], *Haonan Wang*[1], *Qing Ye*[2], *Jiancheng Lv*[2,*] *and Kazunari Matsuda*[1,*]

[1]Institute of Advanced Energy, Kyoto University, Uji, Kyoto 611-0011, Japan

[2]DICALab, School of Computing, Sichuan University, China

[†]These authors contributed equally to this work.

[*]Corresponding authors: Jiancheng Lv, and Kazunari Matsuda

E-mail: lvjiancheng@scu.edu.cn; matsuda@iae.kyoto-u.ac.jp

## Abstract

First-principles structural relaxation and band calculations require substantial computation and expertise. Here we present Perfumer, a large language model-based agentic framework that couples Quantum ESPRESSO with two predictive models. Across tested structures, pre-relaxation with Perfumer-relax model followed by QE refinement yielded a 2.75-fold overall speedup over QE alone. Perfumer-band model combines a periodic graph neural network with a latent Hermitian spectral head returns 12 band energies at any requested wavevector, preserving reciprocal lattice periodicity and time-reversal symmetry without constructing Hamiltonian. Trained on 125,153 self-curated band structures, Perfumer-band model reaches a median per-material mean absolute error of 0.157 eV on 2,607 test materials, including lanthanide, actinide and magnetic compounds, from held-out composition families spanning 85 elements, taking about 97 ms per material on two consumer CPU cores. The agent turns natural-language requests into relaxation-to-band workflows, from a single uploaded structure to automatically constructed new candidates and returns predicted properties.

## Introduction

Advances in computational methods have made first-principles calculations an important tool in modern research, enabling the prediction of physical and chemical properties across a wide range of materials, including insulators, semiconductors and metals. Within all the methods, Density functional theory (DFT) provides a practical framework for first-principles predictions of structural energetics and electronic properties[1,2]. Widely used implementations include Quantum ESPRESSO (QE)[3–5] and the Vienna Ab initio Simulation Package (VASP)[6]. In a typical workflow connecting crystal structures to electronic bands, structural relaxation requires repeated self-consistent evaluations of energies and atomic forces. Band structure calculations then evaluate electronic eigenvalues for the relaxed geometry along selected paths in reciprocal space. These calculations can become costly when screening many candidate materials or modelling large supercells, including moiré structures.

Machine learning (ML) offers routes to accelerating individual stages of these workflows. Neural networks, including graph neural networks (GNNs), predict materials properties and interatomic interactions from atomic structures and local environments[7–10], supported by publicly available datasets such as the Materials Project[11]. For structural relaxation, learned interatomic potentials supply energies and forces for iterative optimization[12–14], whereas DeepRelax predicts relaxed structures from initial configurations[15]. Equivariant architectures such as NequIP and MACE, together with broadly pretrained MACE models, extend these approaches to diverse atomic environments[16–18]. For electronic-structure prediction, DeepH learns DFT Hamiltonian matrix elements in a localized atomic-orbital basis[19], an approach further developed through equivariant networks[20–22]. Other methods, including graph Transformer models, predict band structures directly from crystal structures[23]. These approaches establish complementary ways to reduce the numerical effort required for structural and electronic calculations.

Large language models (LLMs) offer a complementary means of coordinating scientific workflows through agents equipped with computational tools[24–27]. In materials research, multiagent systems have combined atomistic simulations and GNN surrogates for alloy design[28,29]. GENIUS integrates LLMs with domain knowledge to generate QE inputs and uses early execution feedback to validate and repair simulation protocols[30]. MatSciAgent coordinates specialized tools for materials data retrieval, crystal structure generation and molecular dynamics simulations[31]. Such systems translate natural-language requests into computational operations and can reduce the manual effort required to configure and connect scientific tools[32]. Automating calculation setup and execution[33,34] does not, by itself, reduce the cost of the underlying numerical calculations. An integrated relaxation-to-band workflow must therefore address numerical efficiency and task coordination together. It must also transfer structures and calculation settings consistently between stages and distinguish model predictions from quantities refined through first-principles calculations. These requirements motivate combining specialized numerical models with an agent that coordinates their use within a common workflow.

Here we present Perfumer, an LLM-based agentic framework that integrates model-assisted structural relaxation and electronic band prediction with QE. The agent interprets natural-language requests, prepares computational inputs, coordinates tool execution and summarizes numerical outputs. Perfumer-relax model uses a learned interatomic potential to pre-relax atomic structures before QE refinement under specified convergence criteria. Perfumer-band model combines a periodic GNN with a latent Hermitian spectral head that yields selected band energies as eigenvalues at specified reciprocal space points. This representation is trained against reference band energies without reconstructing an orbital-resolved Kohn–Sham Hamiltonian. Its training data cover 91 elements, including all 15 lanthanides and six actinides, as well as magnetic compounds described by spin-polarized bands. The modules can operate independently or sequentially, with the QE-refined geometry passed to Perfumer-band model for electronic band prediction. We assess relaxation efficiency against unassisted QE, evaluate band prediction on held out composition families and illustrate task coordination through natural-language workflow examples.

## Results

### Perfumer framework and training data

Perfumer connects natural-language requests to structural relaxation and electronic band prediction through an LLM coupled to the Perfumer Harness (**Fig. 1**). The LLM interprets tasks and coordinates tool calls, while the numerical calculations are performed by Perfumer-relax model, QE and Perfumer-band model. The harness integrates these specialized tools with general purpose utilities for WebSearch, WebFetch and file access, alongside a scientific database component including prediction setup for commonly used materials. Tool outputs are returned to the LLM to guide subsequent operations and generate the response.

For a combined relaxation-to-band task, Perfumer-relax model first performs model-assisted pre-relaxation. QE then refines the atomic geometry under user defined convergence criteria, and Perfumer-band model predicts selected band energies from the refined structure along the requested reciprocal-space path. The Perfumer-band model predicts 12 electronic bands per spin channel. For nonmetals, these comprise the six highest valence bands and the six lowest conduction bands; for metals, they comprise 12 bands near the Fermi level. The two predictive modules can also operate independently. Users can submit follow-up requests through the same interface to extend or revise the task. The Perfumer agent connects task coordination, first-principles structural refinement and model-based band prediction within a modular workflow.

We also constructed separate training datasets for the two models (**Fig. S1**). For the Perfumer-relax model, we generated a dataset from QE structural relaxation calculations performed on the supercomputer system. Each validated ionic step was stored as a separate record pairing the atomic configuration with the corresponding Cartesian force labels. For the Perfumer-band model, each record associates a crystal structure and a reciprocal space path with 12 selected band energies per spin channel. The curated band dataset contained 130,367 material records and collected data from the Materials Project, JARVIS-DFT, AFLOW and NOMAD[11,35–37]. We grouped band structure records by reduced chemical formula and assigned each group to the training, validation or test set. This allowed us to assess how well Perfumer-band model predicts band energies for compositions unseen during training[38,39]. Dataset construction, target definitions and training procedures are described in **Methods**.

## Architecture and evaluation of Perfumer-relax model

The Perfumer-relax model is based on a MACE interatomic potential fine-tuned on a curated dataset of QE structural relaxation trajectories. The fine-tuned model performs structural pre-relaxation of perturbed or unrelaxed structures, after which QE refines the resulting geometry under the prescribed convergence criteria (**Fig. 2a**). To quantify the benefit, we relaxed each test structure twice on the same compute allocation, with QE alone and with 100 Perfumer-relax steps followed by QE refinement (**Fig. 2b**). For the 553 tested structures, comprising 241 metallic and 207 non-metallic bulk crystals and 92 two-dimensional materials with randomly displaced atoms and 13 displaced 3 × 3 two-dimensional supercells, the total wall time fell from 277.8 h to 101.1 h, a 2.75-fold reduction. The reduction was 3.05-fold for metals, 2.44-fold for non-metals, 2.66-fold for two-dimensional materials and 8.24-fold for two-dimensional supercells.

$Al_6Ir_2$, an eight-atom intermetallic among the metals in **Fig. 2b**, illustrates how pre-relaxation reduces the QE simulation time (**Fig. 2c,d**). QE alone required 419.1 s and 22 force evaluations. With 40, 100 and 200 model steps, the combined pre-relaxation and QE refinement times were 257.7, 54.1 and 53.5 s and the QE force evaluation counts fell to 14, 4 and 4, corresponding to speedups of 1.63, 7.75 and 7.84 (**Fig. 2c**). The final structures agreed with the QE only result to within $4.5 \times 10^{-4}$ Å, with a largest single atom deviation of $6.0 \times 10^{-4}$ Å (**Fig. 2d**).

## Architecture and evaluation of Perfumer-band model

The Perfumer-band model predicts selected electronic band energies from crystal structures, known physical conditions and reciprocal space queries through a periodic latent Hermitian representation (**Fig. 3**). A periodic crystal encoder comprising eight message-passing layers of hidden width 640 and global atom attention produces a crystal representation $g$ and atom representations $\{h_i\}$. These representations support two complementary branches that construct the latent spectral matrix.

The periodic matrix branch decodes structure-dependent coefficients for an expansion in integer Fourier modes. In parallel, an independent query branch combines Fourier features of the reciprocal space coordinate with query-to-atom attention to generate a Hermitian residual. For each requested spin channel, the model constructs:

$$H_\theta(\mathcal{C},\mathbf{q}) = H_{0,\theta}(\mathcal{C}) + \delta_\theta(\mathcal{C})I + \sum_{\mathbf{m}\in\mathcal{M}} \left[T_{\mathbf{m},\theta}(\mathcal{C})e^{i2\pi\mathbf{m}\cdot\mathbf{q}} + T^{\dagger}_{\mathbf{m},\theta}(\mathcal{C})e^{-i2\pi\mathbf{m}\cdot\mathbf{q}}\right] + \gamma R_\theta(\mathcal{C},\mathbf{q}).$$

Here, $\mathcal{C}$ denotes the crystal structure and the declared calculation and band selection conditions, $\mathbf{q}$ is the fractional reciprocal coordinate, and $\mathcal{M}$ is a fixed set of integer frequency vectors. The trainable parameters are denoted by $\theta$, and † denotes the conjugate transpose. Both $H_{0,\theta}$ and the query-dependent residual $R_\theta$ are Hermitian; $\delta_\theta(C)$ is a scalar reference offset predicted by a separate head, which shifts all energies of a channel together, and the learnable scalar gate $\gamma$ controls the residual contribution. The scaled residual $\gamma R_\theta$ corresponds to $\Delta H(\mathbf{q})$ in **Fig. 3**. The selected band energies are obtained from the ordered eigenvalues of the resulting $16 \times 16$ matrix, which carries two buffer bands below and two above the 12 requested bands:

$$\hat{E}_n(\mathcal{C},\mathbf{q}) = s_E\lambda_{n+2}[H_\theta(\mathcal{C},\mathbf{q})], \qquad n = 1,\ldots,12,$$

where $s_E$ is the energy scale determined from the training data, and $\lambda_k$ denotes the $k$th smallest of the 16 eigenvalues; the two lowest and two highest act as buffer bands that are never compared with reference energies. Hermiticity ensures real eigenvalues, and the integer Fourier representations enforce reciprocal lattice periodicity. Because the coefficient matrices are real and the query residual is symmetrized between $\mathbf{q}$ and $-\mathbf{q}$, the predicted energies at $\mathbf{q}$ and $-\mathbf{q}$ are identical by construction, as required by time-reversal symmetry. Each reciprocal space query is evaluated independently at fixed input conditions. The model is trained end to end against reference band energies using an objective that also penalizes errors in dispersion shape, amplitude and adjacent point energy differences. The latent matrix is learned through its eigenvalues. Details of the training objective and output conventions are provided in **Methods**.

On the test set of 2,607 material records spanning 85 elements, the Perfumer-band model achieved a median per-material mean absolute error (MAE) of 0.157 eV. Among the 80 element groups represented by at least ten test records, 77 had median MAEs below 0.30 eV. The median MAEs were 0.127 eV for the 394 records containing lanthanides, 0.223 eV for the 71 records containing actinides and 0.200 eV for the 1,025 spin-polarized records. **Fig. 4a** summarizes the element-resolved results: each cell reports the median MAE for test records containing that element, with $n$ indicating the corresponding number of records. Prediction and export for the full test set took 327.5 s in a single run on two cores of an Intel Core i9-13900H CPU. The median prediction time was 96.9 ms per material. **Fig. S2** presents the corresponding median prediction time for records grouped by element, illustrating variation in processing time across these groups.

We also compared electronic band structures obtained using QE, the Perfumer-band model and Fable 5.2 (max, conversation in Claude Science) for $WSe_2$ by using the same JSON input file. The QE calculation and model prediction are conducted on two cores of the same CPU. The result of band structures and corresponding computation times are compared in **Fig. S3a** and **b**, respectively. Perfumer-band model achieved a band energy MAE of 0.086 eV relative to the QE reference, compared with 0.306 eV for Fable 5.2, corresponding to a 72.0% reduction in MAE. The prediction time was 0.043 s for the Perfumer-band model, compared with 91.3 s for QE and 319 s for Fable 5.2 (whole conversation time). These timings correspond to speedups of approximately 2,100-fold relative to QE and 7,400-fold relative to Fable 5.2. These results demonstrate moderate band structure accuracy combined with rapid prediction and modest computer resource requirements. Moreover, we compared the predicted and DFT reference band structures for selected materials from the test set (**Fig. 4b-c**, **Fig. S4a-d**). For the metal $DyIn_3$ and the non-metal $Ba_2HoO_6Ta$ in **Fig. 4b-c**, Perfumer-band model achieved mean absolute errors of 0.038 and 0.035 eV, respectively. In both cases, the predicted bands closely follow the reference dispersions along the sampled reciprocal space paths. These examples illustrate the model's ability to predict band energies and dispersion features for held-out compositions, complementing the aggregate assessment of generalization across the test set.

**Integrated relaxation-to-band workflow**

The Perfumer agent translated a natural-language request into a coordinated workflow for structural relaxation and electronic band prediction (**Fig. 5**). To facilitate practical use, we also developed a software package that enables users to access and run the integrated workflow (**Fig. S5**). For the $WSe_2$ monolayer example included with the software, the request was "*Please relax the atomic positions of the included $WSe_2$ example structure, calculate its band structure and provide me with a standard QE input.*" The LLM (DeepSeek-V4.1-Flash-1017) planned the task as structure inspection, pre-relaxation with the Perfumer-relax model followed by QE refinement, band prediction with the Perfumer-band model and export of a QE input file. QE converged in two ionic steps to a maximum force of 0.015 eV $Å^{-1}$, and Perfumer-band model predicted 12 bands along Γ–M–K–Γ for the QE-relaxed geometry, with a band gap of 1.48 eV (QE PBE reference: 1.54 eV). The LLM integrated the relaxation log and the predicted bands into a response to the original request within about 5 min, illustrating the connection between task planning, numerical evaluation and result reporting.

The second example extended this workflow to the search for new materials (**Fig. 6**). The request was "*Please identify new materials for chip applications from wurtzite GaN (mp-804) and perform the relax–band workflow for the selected candidates. Relax the candidates one at a time, and predict each candidate's bands right after its own relaxation.*" The LLM imported the wurtzite GaN structure from the Materials Project, generated 12 candidate structures by Ga → Al, Ga → In, Ga → Sc and N → P substitutions and screened them by their predicted bands. It then relaxed five selected structures (AlN, two symmetry-equivalent $AlGaN_2$ structures, $ScGaN_2$ and $Ga_2PN$) one at a time, with Perfumer-relax model pre-relaxation followed by QE refinement, and predicted the band structures of each relaxed geometry with the Perfumer-band model. All five relaxed structures reached maximum forces no more than 0.0091 eV $Å^{-1}$, with predicted band gaps of 1.554 to 3.234 eV; while the $ScGaN_2$ and $Ga_2PN$ compositions do not occur in the band-model dataset. The final output combined summaries with structure files and QE input/output logs, and the whole run took 15 min. This example demonstrated the extension from processing a supplied structure to constructing and evaluating a candidate set within the goal of finding new materials.

## Discussion

We have developed Perfumer, an LLM-based agentic framework that integrates model-assisted structural relaxation, QE refinement and electronic band prediction within a modular computational workflow. The Perfumer-relax and Perfumer-band models address distinct numerical tasks, while the LLM coordinates input preparation, tool execution and result reporting. This organization connects natural language interaction with specialized predictive modules and established first-principles software.

Across 553 structures with randomly displaced atoms, model-assisted pre-relaxation followed by QE refinement reduced the total relaxation time by a factor of 2.75 relative to unassisted QE, counting structures on which both routes reached the same final geometry. On 2,607 test materials from held-out composition families, Perfumer-band model achieved a median per-material mean absolute error of 0.157 eV and took a median of 96.9 ms per material on two CPU cores. These materials span 85 elements, including all 15 lanthanides and six actinides, and 1,025 of them are spin-polarized, most of which are magnetic; the median errors were 0.127 eV for records containing lanthanides, 0.223 eV for those containing actinides and 0.200 eV for the spin-polarized records. These evaluations quantify the relaxation efficiency and band prediction accuracy of the two numerical predictive models.

The integrated workflow passes refined structures between modules and consolidates numerical results and calculation files into a response to the user's request. By coupling material searching and candidate generation with first-principles structural refinement and band prediction, Perfumer supports automated exploration of new material candidates and evaluation of their electronic properties. Future work could quantify how residual structural differences propagate into band predictions and extend the framework to additional electronic properties. Together with systematic validation of the complete workflow, these developments could broaden Perfumer's role in automated materials exploration and discovery.

## Methods

### Training datasets and setup for Perfume-relax model

The relaxation dataset was generated from QE calculations. Each validated ionic step was stored as a separate record linking the atomic configuration to QE Cartesian atomic forces. The curated dataset contained 14,971 ionic configurations covering 62 elements. The retained configurations comprised 12,992 records for three-dimensional systems and 1,979 for two-dimensional systems. For the speed tests, each structure was relaxed twice on the same eight-core allocation of one node of a cluster: with QE (Quantum ESPRESSO 7.5) alone, and with 100 fixed Perfumer-relax steps followed by QE; QE stopped when every Cartesian force component was below $10^{-4}$ Ry $Bohr^{-1}$. The bulk test structures were 500 materials from the relaxation dataset with every Cartesian coordinate displaced by a Gaussian of standard deviation 0.08 Å (mean removed), classified as metallic or non-metallic by the PBE gap of the final self-consistent calculation (threshold 0.1 eV); two-dimensional monolayers and their 3 × 3 supercells were displaced in the same way. Only pairs in which both routes converged to final geometries within 0.05 Å (root-mean-square deviation) were compared. The $Al_6Ir_2$ example in **Fig. 2c,d** is one of the displaced metals of the speed test; its four calculations (QE alone and 40, 100 or 200 model steps followed by QE) ran one after another in a single job on one node.

**Training datasets and setup for Perfumer-band model**

The band dataset incorporated publicly available electronic band structure records and their associated crystal structures from the Materials Project, JARVIS-DFT, AFLOW and the NOMAD Repository, containing 130,367 records. To evaluate generalization to unseen composition families, we partitioned the curated dataset by reduced chemical formula. All material records with the same reduced elemental composition were assigned to a single family, irrespective of differences in crystal geometry, spin configuration, spin-orbit-coupling settings, basis representation or supercell size. Whole families were allocated using a fixed random seed, targeting training, validation and test proportions of approximately 96%, 2% and 2%, respectively. The resulting subsets contained 125,153 training records, 2,607 validation records and 2,607 test records.

Model parameters were initialized randomly and optimized exclusively using training records, including those used for the auxiliary structural objectives. The validation set was used for learning rate scheduling and checkpoint selection according to the predefined validation objective. The test set was excluded from parameter optimization and checkpoint selection. The band model has 45.5 million trainable parameters. It was trained from random initialization with AdamW (peak learning rate $2 \times 10^{-4}$ after 1,000 warm-up updates, weight decay $10^{-4}$, gradient-norm clipping at 1.0) on batches of 32 materials in single precision on one NVIDIA A100 (80 GB) GPU. The learning rate was halved whenever the validation objective had not improved by more than 0.2% for three epochs, down to a floor of $3 \times 10^{-6}$. Model was trained for 200 epochs (782,400 updates).

We trained the Perfumer-band model using a composite objective that combines band energy errors with penalties on dispersion shape, amplitude and adjacent-point energy increments. Each material record contributed equally to the batch objective,

$$\mathcal{L}_{\mathrm{band}}(t) = \frac{1}{|\mathcal{B}|} \sum_{a \in \mathcal{B}} \left[ \ell_E^{(a)} + \alpha(t) \left( 0.25\, \ell_{\mathrm{shape}}^{(a)} + 0.10\, \ell_{\mathrm{RMS}}^{(a)} + 0.10\, \ell_{\mathrm{edge}}^{(a)} \right) \right],$$

where $\mathcal{B}$ denotes a batch of material records and $\alpha(t) = \min(t/1000,1)$, with $t$ counting completed optimization updates. All four terms are dimensionless. More detailed loss setup is illustrated in the **Supplementary Note 1**.

## Data availability

The trained weights of the Perfumer-band and Perfumer-relax models are available in the NewtonLi233/perfumer-agent GitHub repository, together with the source code (Code availability) and the evaluation data generated in this study: the test-set band predictions, the prediction time results, the relaxation speed test results and the agent session logs for the examples in **Figs. 5** and **6**. The band records used for training and evaluation were drawn from the publicly available Materials Project, JARVIS-DFT, AFLOW and NOMAD databases (Methods). The relaxation speed tests and the $Al_6Ir_2$ example were performed using Quantum ESPRESSO 7.5. Both agent sessions were run with DeepSeek-V4.1-Flash-1017 through the official DeepSeek API, and the GaN parent structure was retrieved through the Materials Project API.

## Code availability

The source code for Perfumer is hosted in the NewtonLi233/perfumer-agent GitHub repository, which includes the agent harness, numerical tool interfaces, installation instructions and dependency specifications. The source implementation requires Node.js 22.19.0 or later. The agent harness uses @earendil-works/pi-agent-core v0.85.1, @earendil-works/pi-ai v0.85.1 and Zod v4.6.5. The numerical runtime bundled with Perfumer 1.6.1 comprises Python 3.12.10, PyTorch 2.5.1[40], mace-torch 0.3.16[17], ASE 3.29.0[41], NumPy 2.5.4[42], SciPy 1.18.1[43] and spglib 2.7.0[44]. Perfumer-band was trained in a separate environment using Python 3.12.12, PyTorch 2.8.0 (CUDA 12.6), NumPy 2.2.6, SciPy 1.15.3 and ASE 3.26.0.

The harness accesses DeepSeek models through the official DeepSeek API via an OpenAI-compatible interface. Web search is implemented through the Gemini API using Gemini 3.8 Flash (gemini-3.8-flash) with Grounding with Google Search. Crystal structures are queried and retrieved through the Materials Project API. The harness also supports compatible locally hosted LLMs, including Qwen models, and hosted APIs from OpenAI and Anthropic.

## Acknowledgements

This work was supported by the National Natural Science Foundation of China through the Basic Research Scheme for PhD Students (Grant No. 625B2122) and the National Major Scientific Instrument and Equipment Development Project (Grant No. 62427820); JSPS KAKENHI (Grant Nos. JP20H05664, JP21H05232, JP21H05235, JP22K18986, JP23H05469, JP25H00417, JP25K24710 and 26K21849); and the JST CREST Program (Grant No. JPMJCR24A5).

## Author information

### Authors and affiliations

**Institute of Advanced Energy, Kyoto University, Uji, Kyoto 611-0011, Japan**

Zhiwei Li, Yikai Yang, Haonan Wang and Kazunari Matsuda.

**DICALab, School of Computing, Sichuan University, China**

Yuhao Zhou, Qing Ye and Jiancheng Lv.

### Author contributions

Z.L., Y.Z., J.L. and K.M. designed the overall workflow and methodology. Z.L., Y.Y. and H.W. developed the software. Z.L. and Y.Z. trained the models and analyzed the results. Z.L., Y.Z. and Q.Y. discussed the results. J.L. and K.M. jointly supervised the project. All authors contributed to manuscript preparation. Z.L., Y.Z., Y.Y., J.L. and K.M. reviewed and edited the manuscript. Z.L. and Y.Z. contributed equally to this work.

### Corresponding authors

Correspondence to Jiancheng Lv and Kazunari Matsuda.

## Ethics declarations

### Competing interests

The authors declare that they have no competing interests.

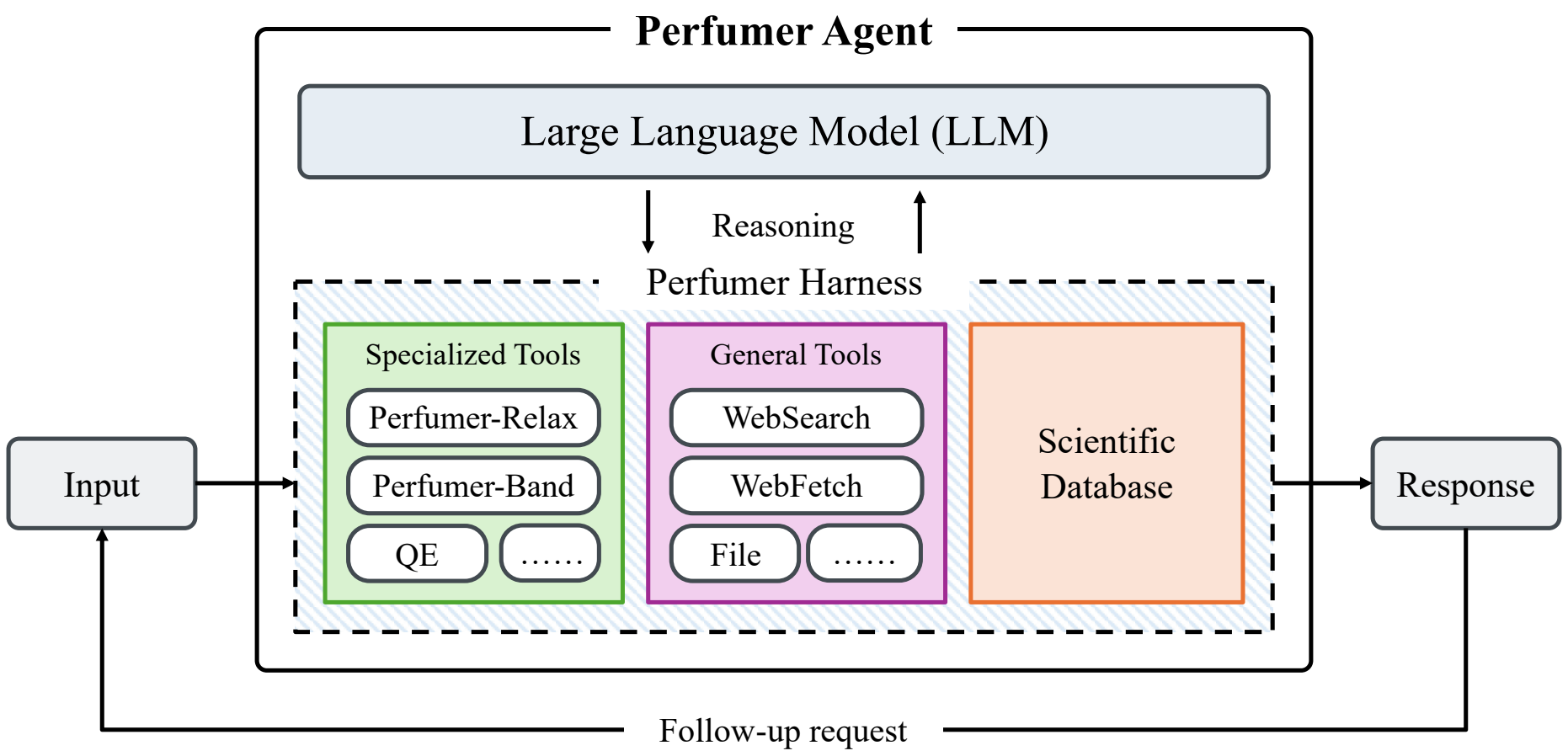

**Fig. 1 | Perfumer agent separates language-based orchestration from numerical prediction.** A single LLM interacts with the Perfumer Harness, which organizes specialized tools for structural relaxation (Perfumer-Relax), Quantum ESPRESSO calculations (QE) and band prediction (Perfumer-Band), alongside general purpose tools (WebSearch, WebFetch, and File) and an in-build database. User requests enter through the Input interface, and results are returned through the Response interface. The follow up request loop allows subsequent requests to re-enter the workflow. The diagram summarizes the system organization and the interfaces between language-based reasoning, numerical tools and user interaction.

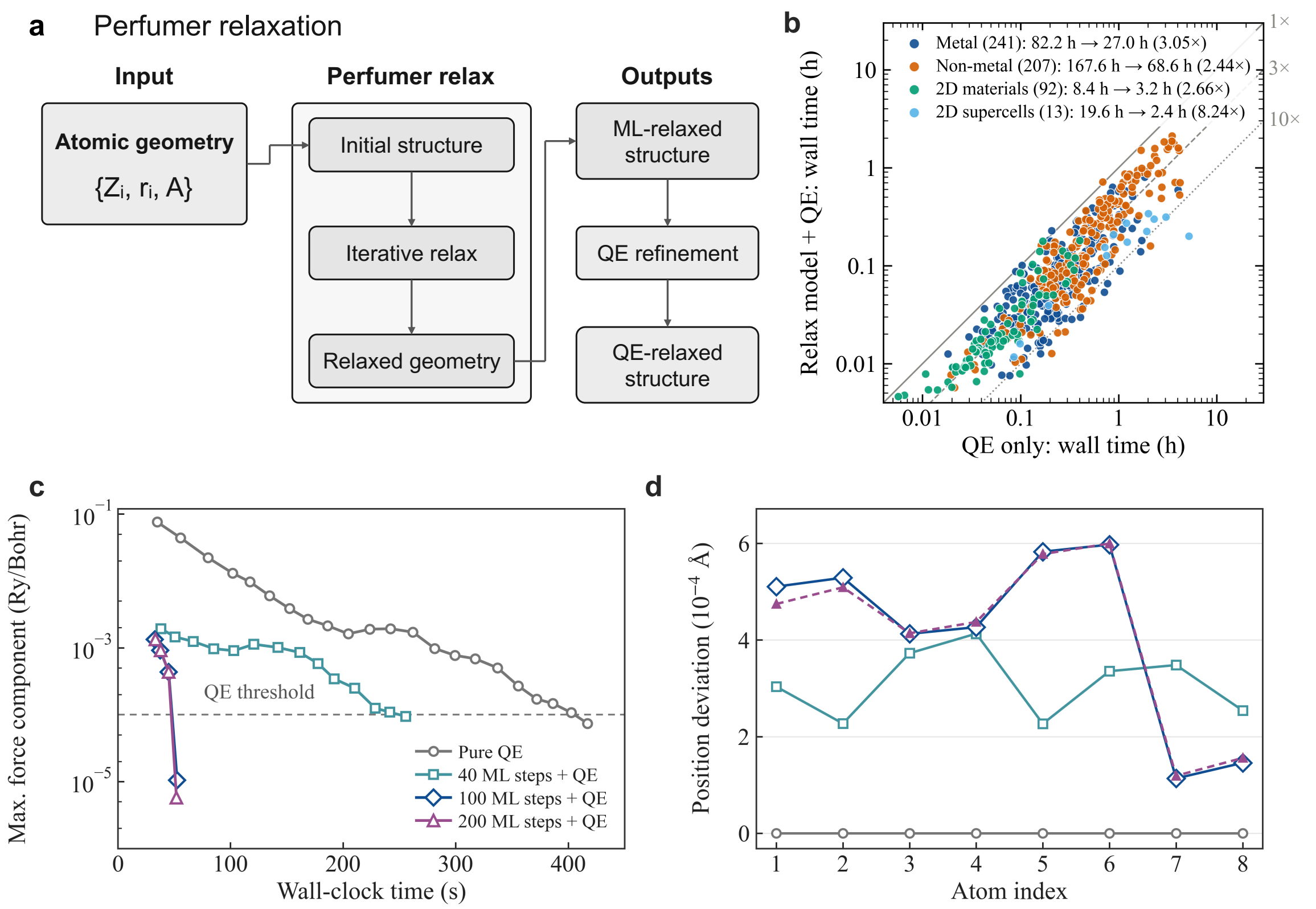

**Fig. 2 | Model-assisted relaxation reduces QE work while retaining close agreement with the QE-only reference endpoint. a**, Workflow from the initial atomic geometry to the QE-refined structure. **b**, Wall time of relaxation with QE alone versus Perfumer-relax model pre-relaxation followed by QE refinement for 553 structures: 241 metals and 207 non-metals (bulk crystals with randomly displaced atoms), 92 two-dimensional materials and 13 two-dimensional supercells. Solid, dashed and dotted lines mark 1×, 3× and 10× speedups; the legend gives the total wall time of each group before and after and their ratio. **c**, Maximum force component versus wall-clock time for the relaxation of $Al_6Ir_2$ with QE alone and with 40, 100 or 200 Perfumer-relax steps followed by QE refinement. Each point is one QE force evaluation; the dashed line marks the QE convergence threshold of $10^{-4}$ Ry $Bohr^{-1}$. **d**, Deviation of the final atomic positions from the QE-only result for the same calculations.

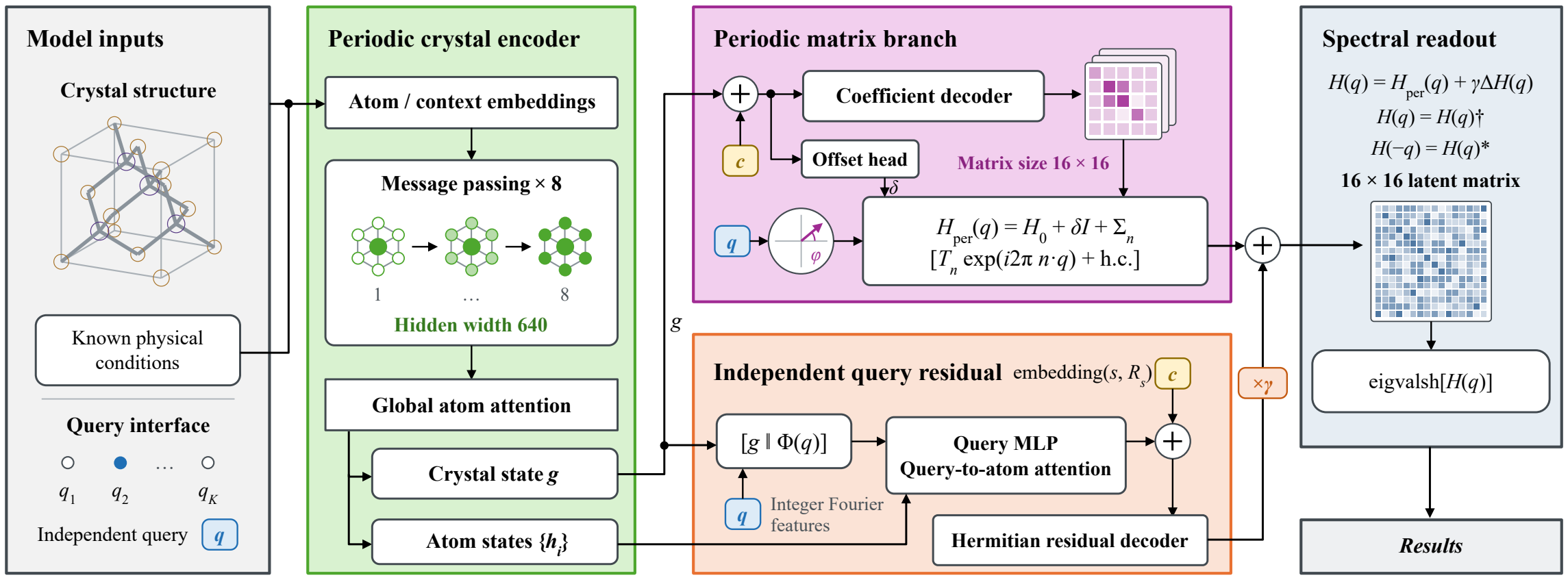

**Fig. 3 | Periodic latent Hermitian architecture of Perfumer-band model.** Crystal structures and known physical conditions are processed by a periodic crystal encoder comprising eight message passing layers of hidden width 640 and global atom attention, yielding a crystal representation $g$ and atom representations $\{h_i\}$. A coefficient decoder constructs the periodic matrix $H_{\text{per}}(\mathbf{q})$ through an expansion in integer Fourier modes (22 modes); a reference-offset head predicts a scalar $\delta$ that shifts all levels through $\delta I$. In parallel, an independent query branch combines $g$, Fourier features $\Phi(\mathbf{q})$ and query-to-atom attention to generate a Hermitian residual $\Delta H(\mathbf{q})$, scaled by a learnable gate. The two contributions form the $16 \times 16$ latent Hermitian matrix $H(\mathbf{q}) = H_{\text{per}}(\mathbf{q}) + \gamma \Delta H(\mathbf{q})$, whose 3rd to 14th ordered eigenvalues, after energy rescaling, provide 12 predicted band energies per spin channel. The two lowest and two highest eigenvalues act as buffer bands. The conditioning vector $c$ encodes the requested spin channel $s$ and band rank indices $R_s$; h.c. denotes the Hermitian conjugate. Reciprocal space queries are evaluated independently. Real coefficient matrices and a residual symmetrized between $\mathbf{q}$ and $-\mathbf{q}$ make the predicted energies at $\mathbf{q}$ and $-\mathbf{q}$ identical. The latent matrix is trained against reference band energies without reconstructing an orbital-resolved Kohn–Sham Hamiltonian.

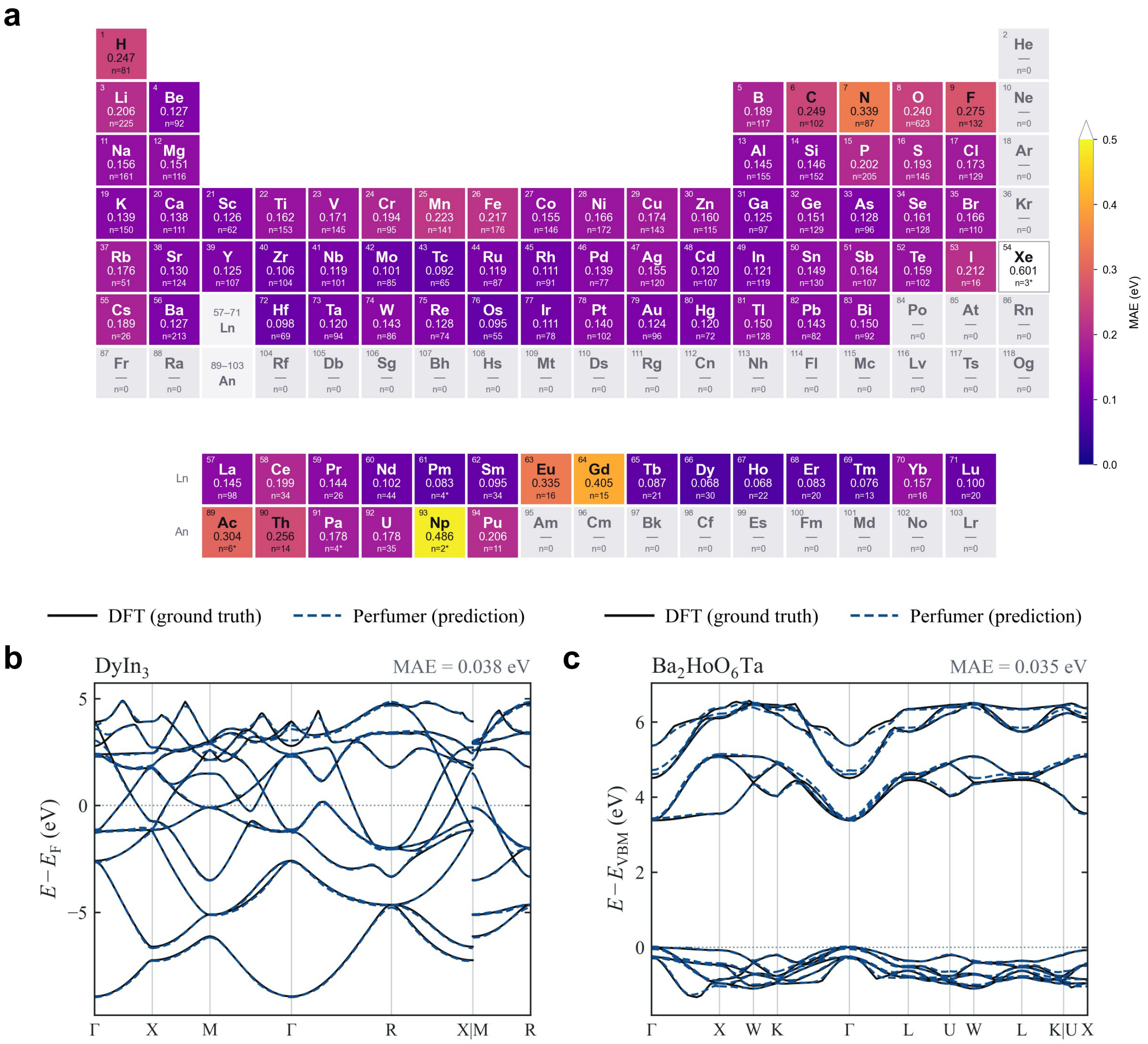

**Fig. 4 | Element-resolved evaluation and selected band structure predictions of Perfumer-band. a**, Band energy errors across elements in the held-out test set of 2,607 material records. Cell colours and numerical values indicate the median per-material MAE for records containing each element, and $n$ denotes the corresponding number of records. Each record contributes to every element it contains. Asterisks indicate $n < 10$, while grey element cells indicate no test records. White cells mark element medians above the colour-scale maximum of 0.5 eV (Xe, 0.601 eV, n = 3). **b**,**c**, DFT reference bands (black solid lines) and Perfumer-band model predictions (blue dashed lines) for the metal $DyIn_3$ (**b**) and the non-metal $Ba_2HoO_6Ta$ (**c**), with MAEs of 0.038 and 0.035 eV, respectively. Energies are referenced to the Fermi level, $E_F$, in **b** and to the valence-band maximum, $E_{VBM}$, in **c**.

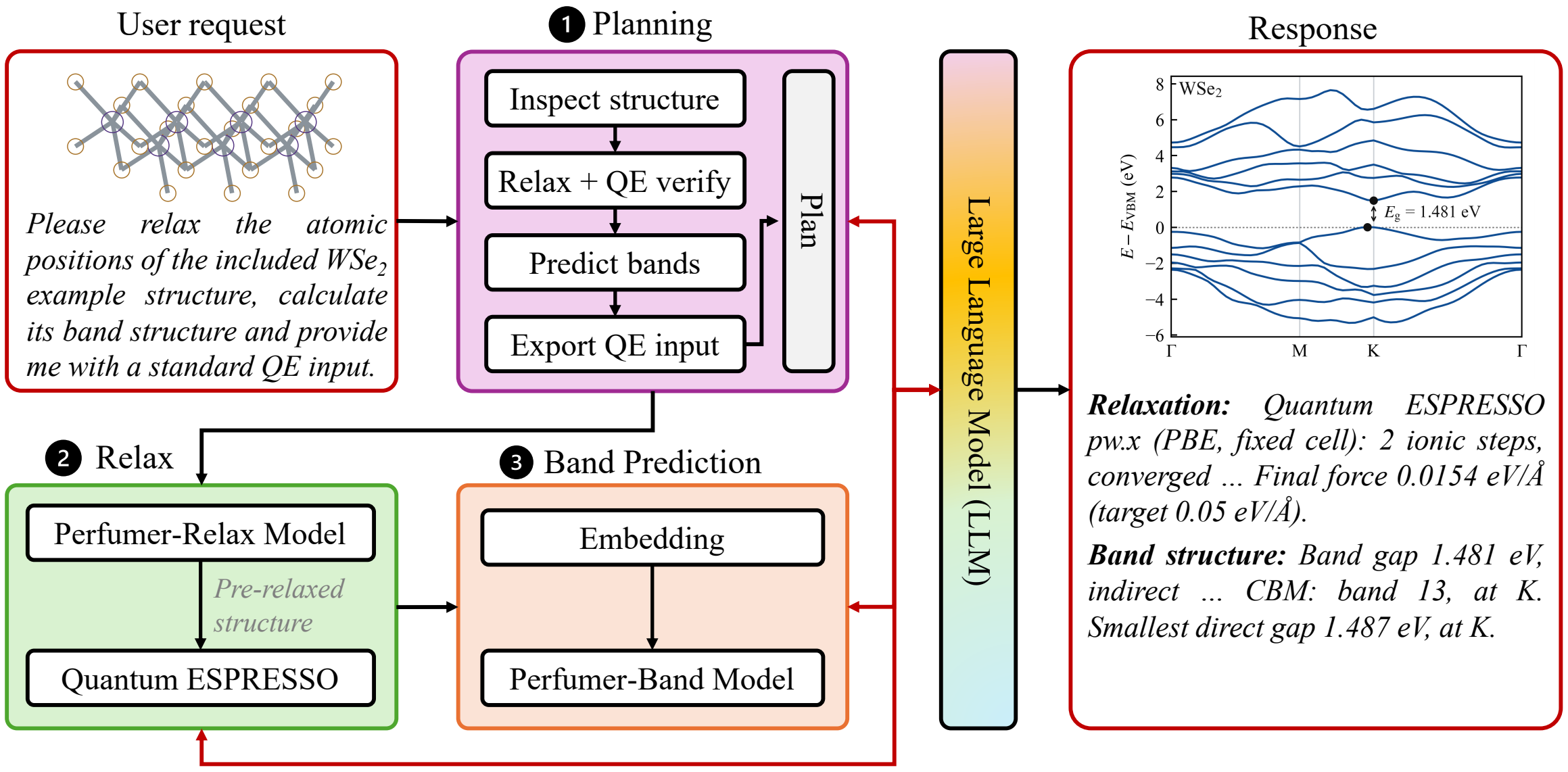

**Fig. 5 | LLM-coordinated structural relaxation and band prediction for monolayer $WSe_2$.** A natural-language request is translated into a plan comprising structure inspection, model-assisted pre-relaxation, QE refinement[45], band prediction and QE input export. The LLM coordinates the numerical tools: Perfumer-relax model provides the starting geometry for QE refinement, and the refined structure is encoded and passed to Perfumer-band model. Numerical outputs are returned to the LLM and assembled into a response summarizing structural relaxation and predicted electronic properties. The QE relaxation was run at fixed lattice vectors and converged in two ionic steps to a maximum force of 0.015 eV Å$^{-1}$. The displayed bands and the reported band structure are model predictions for the QE refined geometry (QE PBE reference: direct gap of 1.54 eV at K). The run used Perfumer 1.6.1 with DeepSeek-V4.1-Flash-1017 as the LLM.

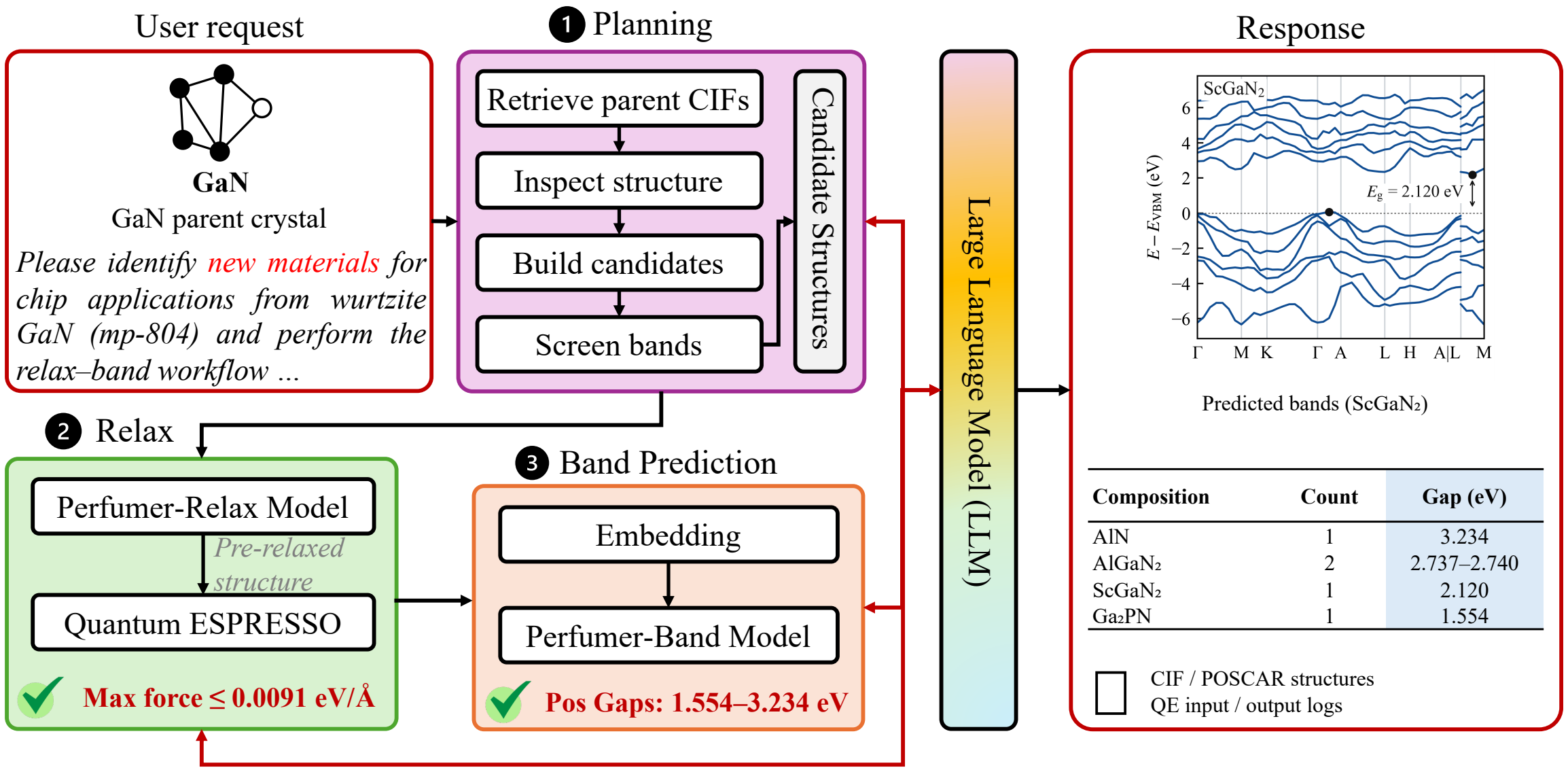

**Fig. 6 | Automated construction and evaluation of candidate materials with Perfumer.** In response to a natural-language request to identify new materials for chip applications from wurtzite GaN, the Perfumer agent imports the GaN parent structure (mp-804) from the Materials Project, inspects it, constructs 12 candidate structures through Ga → Al, Ga → In, Ga → Sc and N → P substitutions and screens their predicted bands. Five selected structures then undergo model-assisted pre-relaxation, QE refinement and Perfumer-band model prediction under LLM coordination. The relaxed structures have final maximum forces no greater than 0.0091 eV Å$^{-1}$ and predicted band gaps of 1.554–3.234 eV. The output includes the predicted bands of $ScGaN_2$, an overall summary, CIF/POSCAR structures and QE input/output logs. Counts indicate the number of evaluated structures for each composition. The displayed bands and gaps are model predictions for the QE-refined geometries.

# Supporting Information for

# Perfumer: A Large Language Model-based Agentic Framework Accelerating Structural Relaxation and Electronic Band Prediction for First-Principle Calculation

*Zhiwei Li[1,†], Yuhao Zhou[2,†], Yikai Yang[1], Haonan Wang[1], Qing Ye[2],*

*Jiancheng Lv[2,*] and Kazunari Matsuda[1,*]*

[1]Institute of Advanced Energy, Kyoto University, Uji, Kyoto 611-0011, Japan

[2]DICALab, School of Computing, Sichuan University, China

[†]These authors contributed equally to this work.

[*]Corresponding authors: Jiancheng Lv, and Kazunari Matsuda

E-mail: lvjiancheng@scu.edu.cn; matsuda@iae.kyoto-u.ac.jp

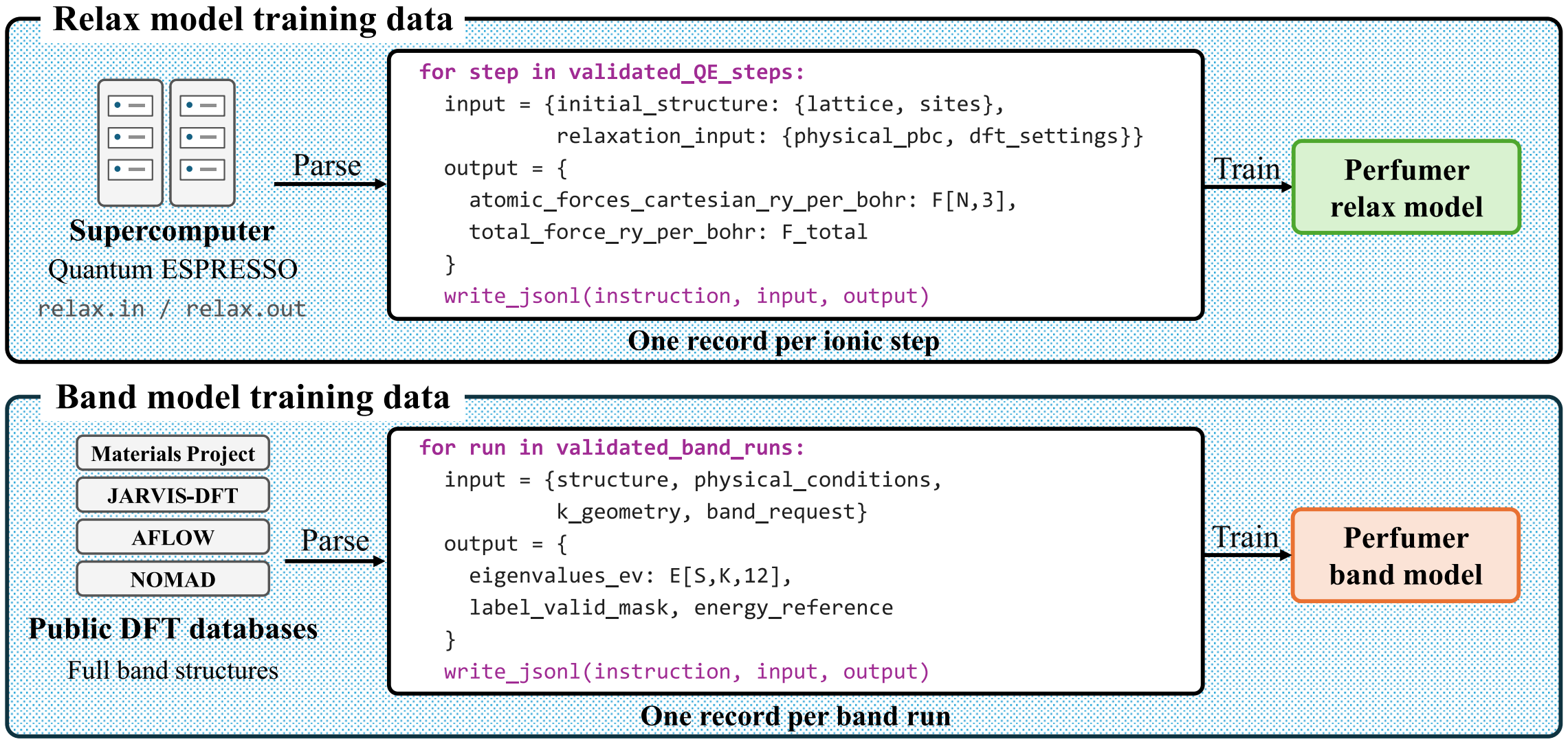

**Fig. S1 | Distinct reference sources define the two training pipelines.** The upper pipeline parses QE relax.in/relax.out files into one force supervised record per ionic step, using the structure, physical periodicity and DFT settings as inputs. The lower pipeline parses band structure records from the Materials Project, JARVIS-DFT, AFLOW and the NOMAD Repository (130,367 records). Energies are referenced to the VBM for eligible nonmetals and to the Fermi level for metals.

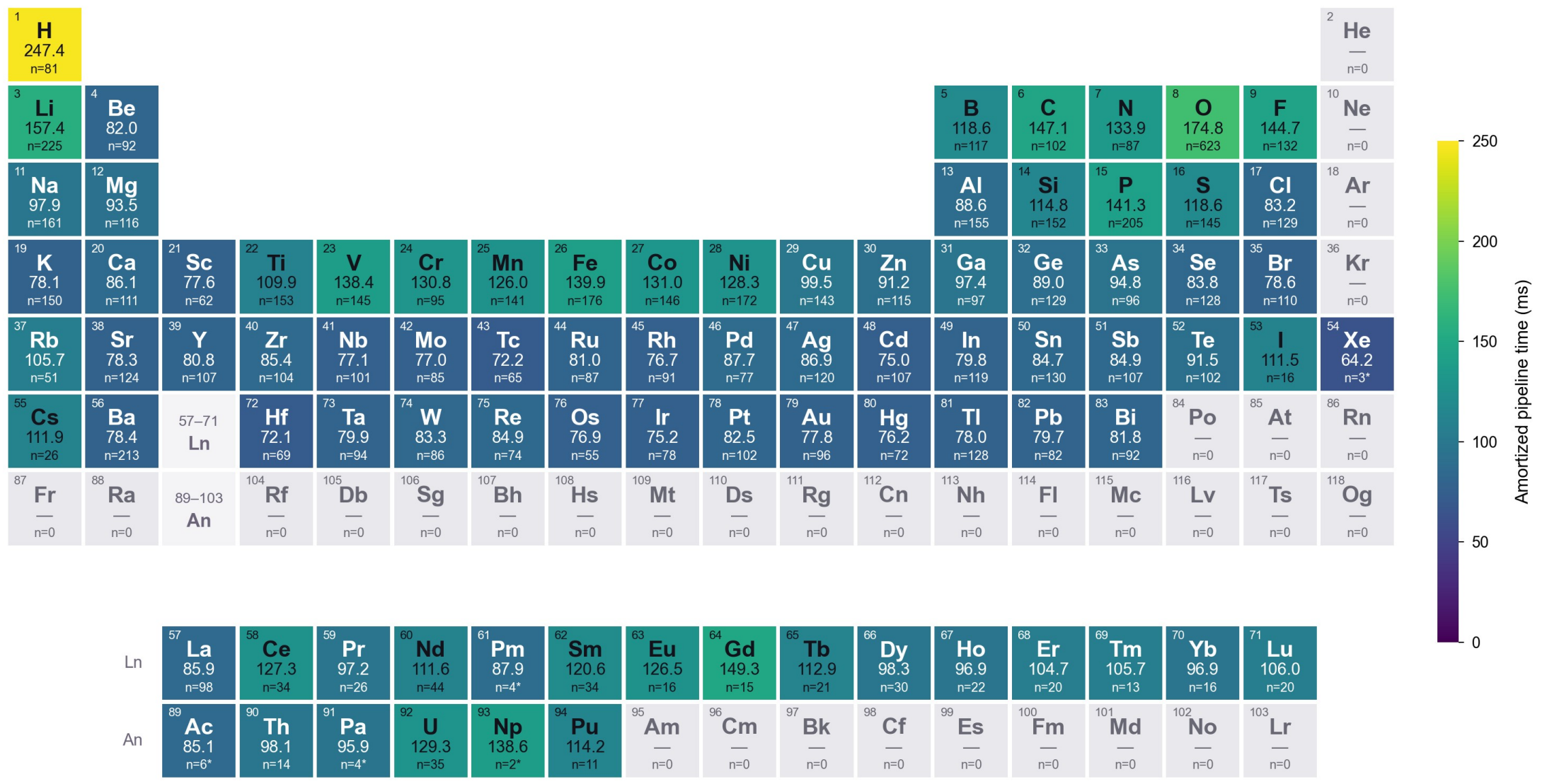

**Fig. S2 | Element resolved timings for the Perfumer-band model prediction time.** Cell colours and numerical values indicate the median prediction time, in milliseconds, for the 2,607 test materials containing each element, measured in a single run on two cores of an Intel Core i9-13900H CPU. The value $n$ denotes the number of test records containing each element; records containing multiple elements contribute to each corresponding group. Asterisks indicate $n < 10$, while grey element cells indicate no test records.

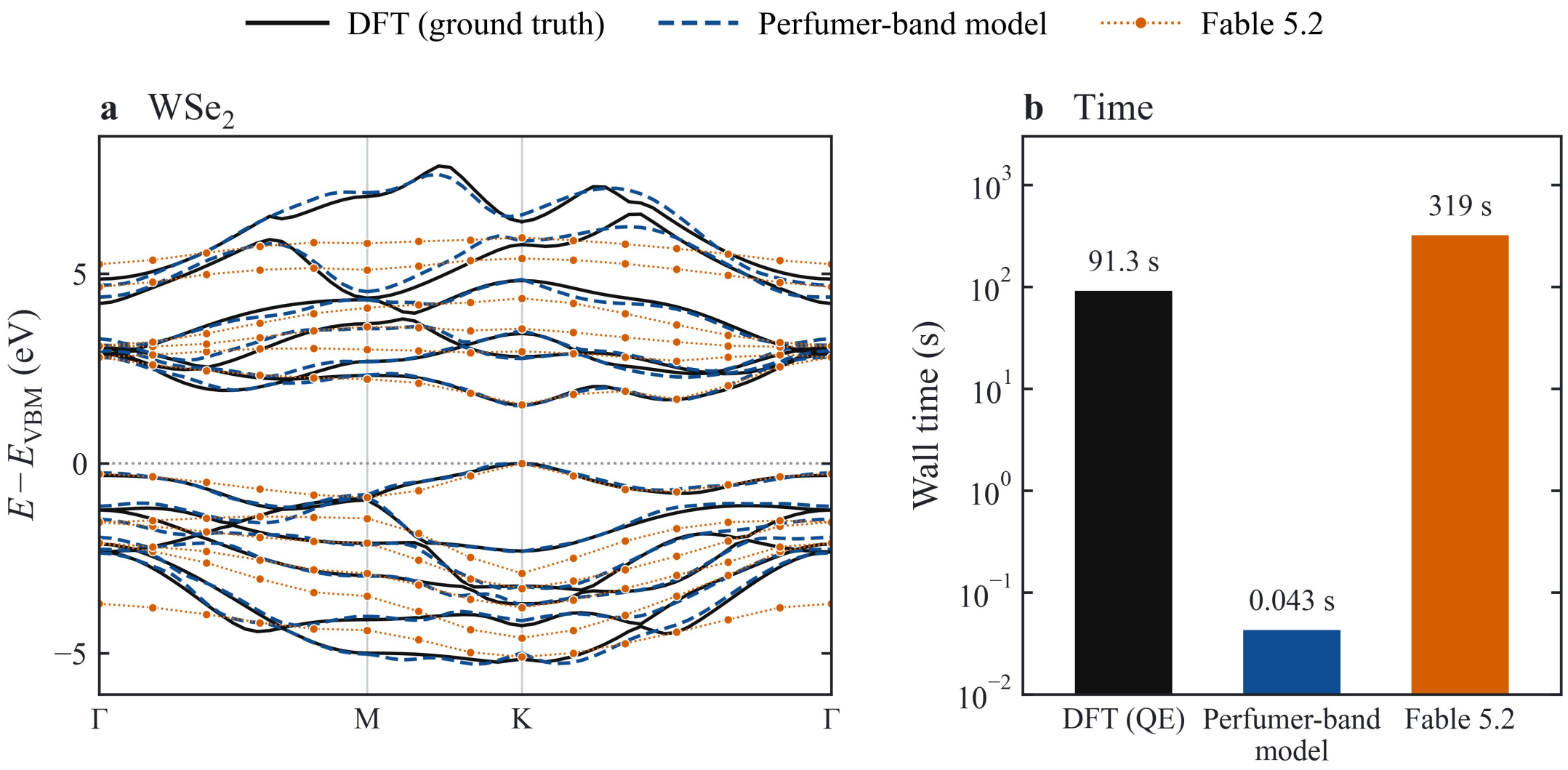

**Fig. S3 | Comparison of electronic-band predictions and computation times for $WSe_2$.** **a**, QE reference bands (black solid lines) and predictions from the Perfumer-band model (blue dashed lines) and Fable 5.2 (orange dotted lines with markers) for a $WSe_2$ structure. The MAEs across the 12 selected bands are 0.086 eV and 0.306 eV for Perfumer-band model and Fable 5.2, respectively. Energies are referenced to the valence-band maximum, $E_{\mathrm{VBM}}$. **b**, Corresponding wall times on a logarithmic scale. QE calculations and Perfumer-band model predictions were performed on an Intel Core i9-13900H processor using two cores; the Fable 5.2 time is that of the whole conversation.

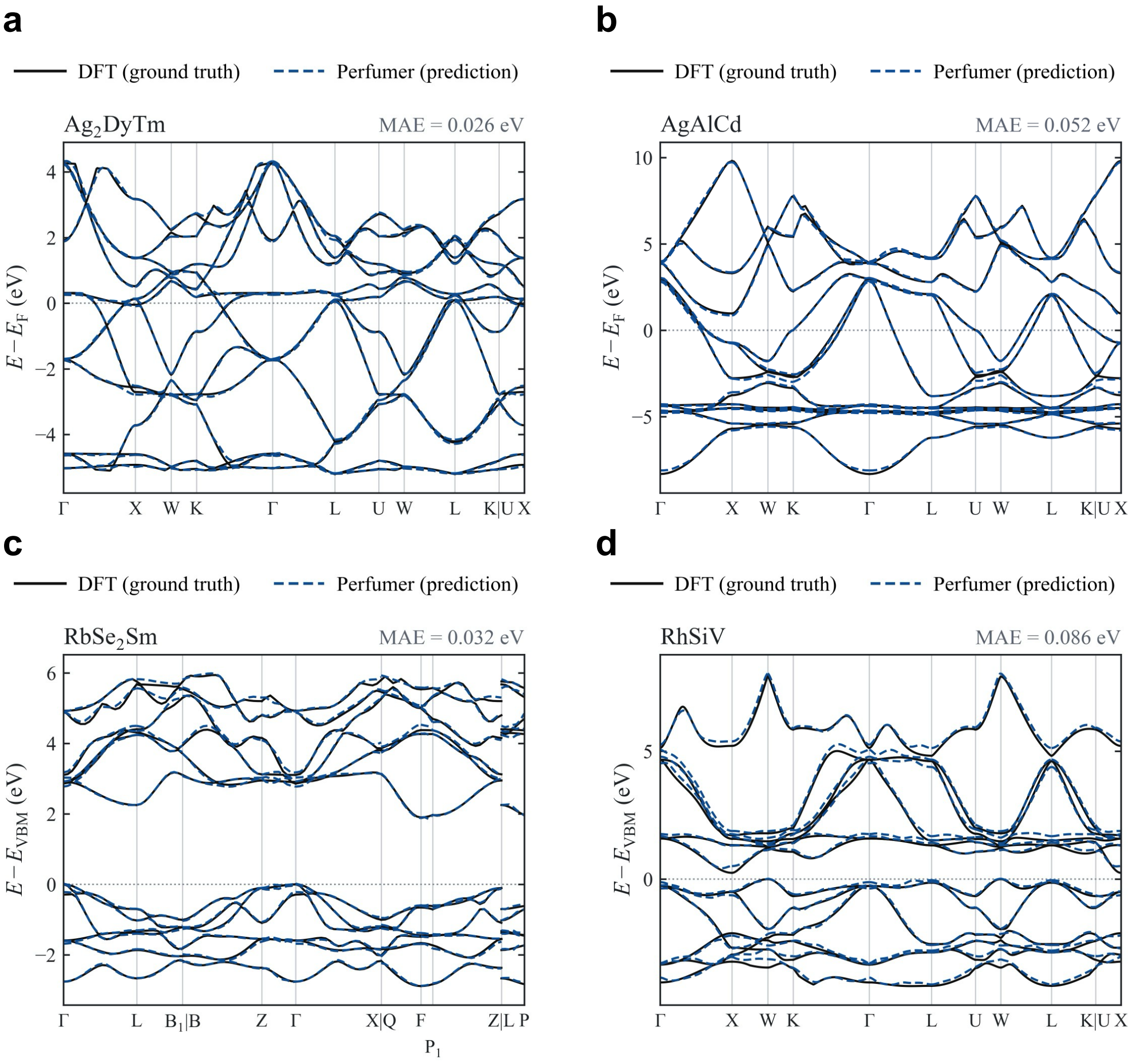

**Fig. S4 | Additional examples of band structure predictions. a–d**, Comparisons between DFT reference bands (black solid lines) and Perfumer-band model predictions (blue dashed lines) for four further test materials, two metals (**a,b**) and two non-metals (**c,d**). The mean absolute error is indicated in each panel. Energies are referenced to the Fermi level in **a,b** and to the valence-band maximum in **c,d**.

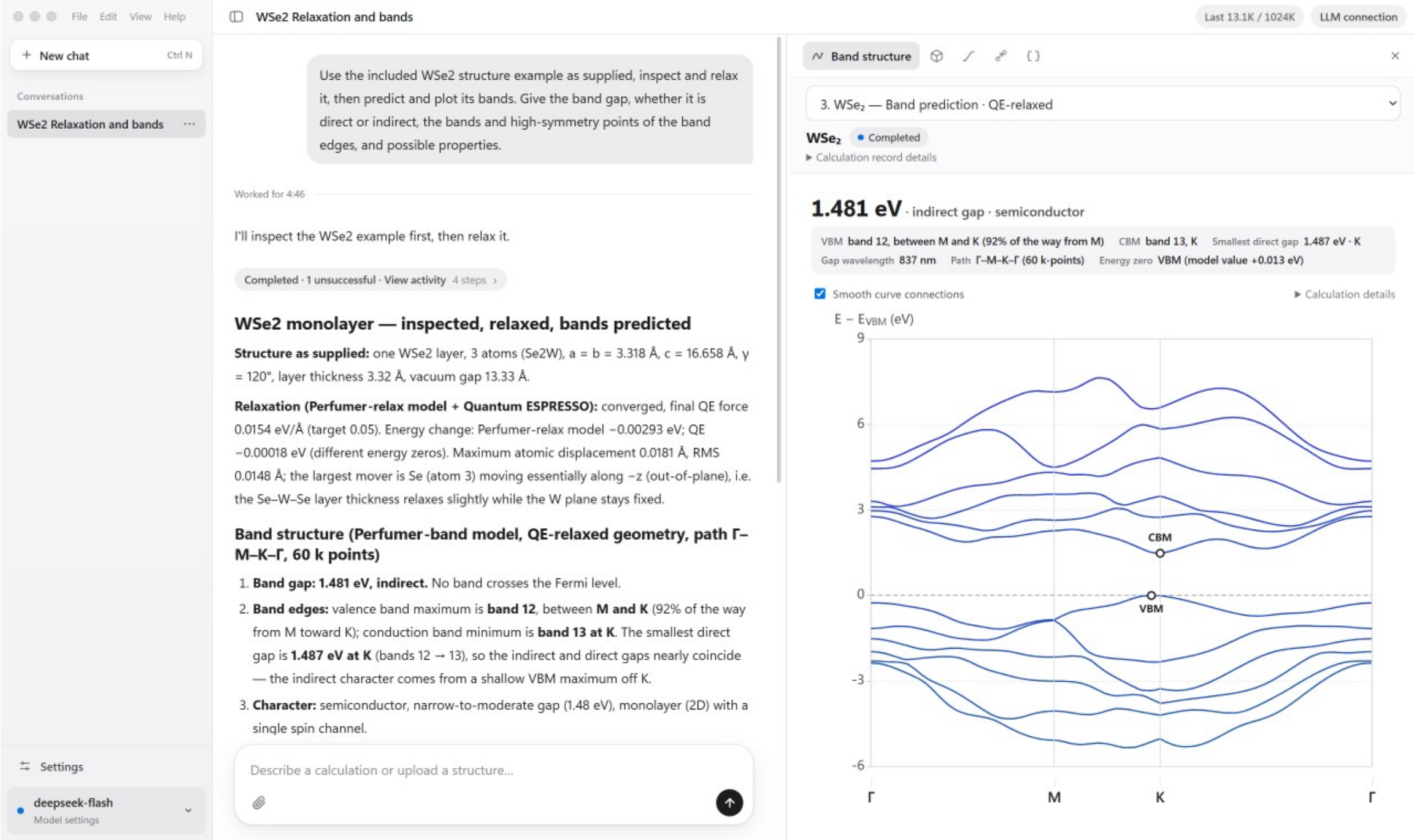

**Fig. S5 | User interface of the Perfumer software.** Screenshot showing natural-language task submission, calculation summaries and visualization of predicted electronic bands for a $WSe_2$ example.

**Supplementary Note 1: Detailed loss setup of Perfumer-band model**

For material $a$, we denote the predicted and reference energies by $\hat{E}_{abj}$ and $E_{abj}$, respectively, where $b$ identifies a fixed energy rank within a physical spin channel and $j$ indexes the sampled $k$-points. Let $\mathcal{K}_{ab}$ contain the valid query indices, $N_{ab} = |\mathcal{K}_{ab}|$, and $\mathcal{T}_a$ contain the tracks with at least one valid label. The energy term was defined as

$$\ell_E^{(a)} = \frac{\sum_{b\in\mathcal{T}_a}\sum_{j\in\mathcal{K}_{ab}}\left|\hat{E}_{abj} - E_{abj}\right|}{s_E \sum_{b\in\mathcal{T}_a} N_{ab}}, \qquad s_E = 0.10 \text{ eV}.$$

This term retained the declared energy reference without fitting an additional energy offset.

To supervise dispersion shape separately from its mean energy, each predicted and reference track was centered over its valid points along the entire declared path. We define $\delta E_{abj} = E_{abj} - \overline{E}_{ab}$, where $\overline{E}_{ab} = N_{ab}^{-1}\sum_{j\in\mathcal{K}_{ab}} E_{abj}$, and define $\delta\hat{E}$ analogously. The shape term was

$$\ell_{\text{shape}}^{(a)} = \frac{1}{|\mathcal{T}_a|}\sum_{b\in\mathcal{T}_a}\frac{1}{N_{ab}}\sum_{j\in\mathcal{K}_{ab}} \rho\left(\frac{\delta\hat{E}_{abj} - \delta E_{abj}}{s_{\text{shape}}}\right), \qquad s_{\text{shape}} = 0.05 \text{ eV},$$

where $\rho(x) = x^2/2$ for $|x| \le 1$ and $\rho(x) = |x| - 1/2$ otherwise. This averaging gives equal weight to each nonempty track within a material.

Dispersion amplitudes were compared using a regularized logarithmic root-mean-square term. With $v_{ab} = N_{ab}^{-1}\sum_j \delta E_{abj}^2$ and the corresponding predicted variance $\hat{v}_{ab}$, we used

$$\ell_{\text{RMS}}^{(a)} = \frac{1}{|\mathcal{R}_a|}\sum_{b\in\mathcal{R}_a} \rho\left[\frac{1}{2s_{\log}}\log\left(\frac{\hat{v}_{ab} + \epsilon^2}{v_{ab} + \epsilon^2}\right)\right],$$

where $s_{\log} = 0.30$ and $\epsilon = 0.001$ eV. The eligible set $\mathcal{R}_a$ contained tracks with at least two valid points and a reference centered RMS of at least 0.01 eV.

Local energy variations were supervised through adjacent-point increments,

$$\ell_{\text{edge}}^{(a)} = \frac{1}{|\mathcal{A}_a|}\sum_{(b,j,k)\in\mathcal{A}_a} \rho\left[\frac{\left(\hat{E}_{abk} - \hat{E}_{abj}\right) - \left(E_{abk} - E_{abj}\right)}{s_{\text{edge}}}\right], \qquad s_{\text{edge}} = 0.05 \text{ eV}.$$

The set $\mathcal{A}_a$ included consecutive query pairs within the same path segment, with valid labels at both endpoints and a Cartesian reciprocal-space separation exceeding $10^{-7}$ Å$^{-1}$. For the evaluated rank, edges touching a reference state within 0.02 eV of another valid rank in the same spin channel, or an explicitly flagged unresolved crossing, were excluded. The increments were not divided by the $k$-point separation and therefore did not constitute velocity supervision. Terms with no eligible entries contributed zero.

Structural auxiliary objectives were evaluated every fourth update on two distinct structures sampled from the deduplicated training pool. The complete optimization objective was

$$\mathcal{L}_{\text{train}}(t) = \mathcal{L}_{\text{band}}(t) + 0.02\,\mathbf{1}_{t \bmod 4 = 0}\,\mathcal{L}_{\text{struct}},$$
$$\mathcal{L}_{\text{struct}} = \mathcal{L}_{\text{species}} + 0.25\,\mathcal{L}_{\text{denoise}} + 0.10\,\mathcal{L}_{\text{consistency}} + 0.10\,\mathcal{L}_{\text{variance}}.$$

These objectives comprised masked-species reconstruction, distance denoising, consistency between normalized projected structure representations, and variance regularization. Species were masked with probability $0.15$, with at least one masked site per structure. Denoising used Cartesian Gaussian perturbations with a standard deviation of $0.03$ Å. The auxiliary objectives used structure-only inputs and no electronic labels. Checkpoint selection used the full-validation band objective with $\alpha = 1$; the unnormalized energy MAE in eV was monitored separately.